\documentclass[journal]{IEEEtran}
\usepackage{graphicx}
\usepackage{multicol}
\usepackage{tabularx}
\usepackage{multirow}
\usepackage{epstopdf} %converting to PDF
\usepackage{cite}
\usepackage[table]{xcolor}

\usepackage{threeparttable}
\begin{document}

\title{Harnessing the Potential of Optical Communications for the Metaverse}

\author{Baha Eddine Youcef Belmekki,~\IEEEmembership{Member,~IEEE,}
        Abderrahmen Trichili,~\IEEEmembership{Senior Member,~IEEE,}
    Boon~S.~Ooi,~\IEEEmembership{Fellow,~IEEE, 
    }%  <-this % stops a space
        and~Mohamed-Slim~Alouini,~\IEEEmembership{Fellow,~IEEE,}
\thanks{
B.~E.~Y.~Belmekki, B.~S.~Ooi, and M.~-S.~Alouini are with the Computer, Electrical and Mathematical Sciences $\&$ Engineering in King Abdullah University of Science and Technology, Thuwal, Makkah Province, Saudi Arabia. Email: \{bahaeddine.belmekki, boon.ooi, slim.alouini\}@
kaust.edu.sa. A.~Trichili carried out this work when he was at KAUST, and is currently with the Department of Engineering Science of the University of Oxford. Email: abderrahmen.trichili@kaust.edu.sa.
}
}

\markboth{IEEE Communications Magazine}%
{Baha, Abderrahmen, Prof. Boon, and Prof. Slim}

\maketitle

\thispagestyle{empty}
\pagestyle{empty}
\begin{abstract}
%The Metaverse is a digital world that uses virtual reality (VR) to create an immersive virtual experience.
%However, the Metaverse requires ultra-high-speed connectivity to enable bandwidth-hungry applications. To cater for these requirements, optical communication arises as a key pillar in bringing this paradigm into reality. 
%In this paper, we highlight the potential of optical communications in the Metaverse.
%First, we set forth the \Rev{requirements} and ultra-high data rate requirements for various Metaverse experiences. 
%Then, we put forward the potential of optical communications to achieve these data rate requirements.
%Both optical fiber and optical wireless communication technologies are detailed as well as their current and future expected data rates.
%We also highlight the capability of optical wireless communication for joint communication, localization, and sensing in the access component.
%In addition, we propose a set of configurations, connectivity, and equipment necessary for the Metaverse.
%Finally, we identify a set of research directions that will further improve the use of the optical spectrum and tackle the related challenges to meet the growing needs of the upcoming Metaverse era.

The Metaverse is a digital world that offers an immersive virtual experience. However, the Metaverse applications are bandwidth-hungry and delay-sensitive that require ultrahigh data rates, ultra-low latency, and hyper-intensive computation. To cater for these requirements, optical communication arises as a key pillar in bringing this paradigm into reality. We highlight in this paper the potential of optical communications in the Metaverse. 
First, we set forth Metaverse requirements in terms of capacity and latency; then, we introduce ultra-high data rates requirements for various Metaverse experiences. 
Then, we put forward the potential of optical communications to achieve these data rate requirements in backbone, backhaul, fronthaul, and access segments. Both optical fiber and optical wireless communication (OWC) technologies, as well as their current and future expected data rates, are detailed. In addition, we propose a comprehensive set of configurations, connectivity, and equipment necessary for an immersive Metaverse experience. Finally, we identify a set of key enablers and research directions such as analog neuromorphic optical computing, optical intelligent reflective surfaces (IRS), hollow core fiber (HCF), and terahertz (THz).

%In addition, we propose a set of configurations, connectivity, and equipment necessary for the Metaverse. 
%Finally, we identify a set of key enablers and research directions that will further 
%Increase computationalcapabilities; Analog Neuromorphic Optical Computing
%reduce latecy transmission hollow core fiber, improve th spectral eeffeciency and mitintge  blockage and user  mobility via IRS and wavefront designe, and improve the spectrum uusage bsing using terahetz calongside optical spectgrum.

\end{abstract}

%\begin{IEEEkeywords}
%Metaverse, virtual reality, optical fiber communication, optical wireless communication, radio-over-fiber, LiFi, optical sensing and localization.
%\end{IEEEkeywords}

\IEEEpeerreviewmaketitle

\section{Introduction}

\IEEEPARstart{M}{etaverse}, is what many consider to be the logical continuation of the internet \cite{tang2022roadmap,chang20226g}. Although different definitions of the Metaverse exist, the dominant ones state that users can use the virtual world and participate in different activities such as gaming, work, and socializing \cite{bui2023game}. Therefore, it is expected to be a user-centric next-generation Internet \cite{du2022exploring}.
The Metaverse attracted a lot of attention during the last year, and big tech companies are vying to acquire and shape its development and tap into its highly anticipated profitable markets. Estimation suggests that the Metaverse industry will grow from 22.79 billion USD in 2021 to 996.42 USD billion in 2030 \cite{GlobalData}.
Tech giants such as Apple, Sony, Google, Microsoft, and Facebook are investing heavily in creating their own Metaverse, and other tech companies are lining up to enter the market. To show its commitment to the Metaverse, Facebook rebranded its name to Meta, thus, steering the company in a new direction.

In the Metaverse, users can explore the virtual worlds via virtual reality (VR) and augmented reality (AR) headsets \cite{xu2022full}.
VR and AR systems have become extremely popular with the advent of the Metaverse and by enabling new immersive virtual experiences. The application areas of VR and AR in the Metaverse are not only limited to gaming and entertainment, but also education and training, and healthcare \cite{khan2019visible}, with an even broader gamut of use cases expected in the future.
 
The paradigm of the Metaverse started coming to life thanks to the rapid advancement of VR headsets, haptic technologies, wireless communication networks, and edge computers. However, the stringent requirements on the VR headset, data rate, and latency, stand in the way of a real-time and scalable implementation of the Metaverse \cite{xu2022full}.
A lifelike immersive experience imposes high constraints on the VR headset to mimic human visual perception; and on the high data rates to transmit the tremendous amounts of data required ($>$2~Tbps) \cite{HVR,wang2022meta}.
In addition, the Metaverse requires low end-to-end latency ($<$20~ms) for an immersive experience without causing motion sickness to the user \cite{cuervo2018creating} \cite{saad2019vision,bastug2017toward,bui2023game}.
These stringent high data rates and low latency requirements combined with high-speed real-time data processing place tremendous strain on communication and computational infrastructures. 
The current fifth-generation (5G) technologies, such as mmWave, cannot cater to communication requirements \cite{wang2022meta,khan2019visible}.
Therefore, it is paramount that the sixth generation (6G) wireless systems can fulfill the aforementioned requirements to offer a fully-immersive Metaverse experience \cite{KhanArxiv22}.

Optical communications provide an end-to-end solution, from the backbone, through the backhaul and fronthaul, to the Metaverse end-user using optical fiber and optical wireless communication (OWC).
By leveraging the broad and unlicensed optical spectrum, optical communications will enable unprecedented data rates over optical fiber and free space channels required for the Metaverse.
In addition, indoor OWC links can provide the required ultra-high data rates, and the in-room light signals' confinement offers extra security imposed in Metaverse applications. OWC can also be used for sensing and localization.
Finally, optical computing can provide higher bandwidths and data rates with 
lower latencies and lower power consumption than their electronic counterpart \cite{wu2022analog}.
%Here I will add a couple of selling points
In this article, we first detail the requirements of the Metaverse with the data rates required for the different types of immersion in the Metaverse in Section~\ref{section2}. We then show, in Section~\ref{section3} and Section~\ref{section4}, how the steady increase in optical fiber and wireless networks will fulfill such requirements in the access, fronthaul, and backhaul components. In addition, we detail in Section~\ref{section5} the Metaverse configurations, connectivity settings, and equipment to enable the Metaverse.
We further discuss key enablers and future research directions to harness the full potential of optical communication to satisfy the requirements of the Metaverse in Section~\ref{section6}.
% What is this paper all about ? sell it with style
% We show the Metaverse requirement in terms of data rates
% we sell the optical communication as Optical fiber and wireless solution solutions 
% in the Access: how the optical wireless solution enables the Metaverse stringent data rates requirement
% we highlight the future research directions
\section{Metaverse Requirements, Data Rate Desiderata, and Enabling Technologies}\label{section2}
In this section, we will detail the Metaverse requirements,
the wireless data rates requirements, and the key enabling technologies that can fulfill these requirements.
\begin{table*}[ht!]
\begin{threeparttable}
\centering
\caption{Data rates for a Metaverse experience for different refresh rates, rendering types, and compressing rates.}
\begin{tabular}{|m{75pt}|m{41pt}|m{41pt}|m{41pt}|m{41pt}|m{41pt}|m{41pt}|m{41pt}|m{39pt}|m{39pt}|m{39pt}|}
\hline
Refresh Rate (Eye angular velocity)&\multicolumn{2}{c|}{120 Hz (2°/s)}&\multicolumn{2}{c|}{200 Hz (4°/s)}&\multicolumn{2}{c|}{500 Hz (8°/s)}&\multicolumn{2}{c|}{1800 Hz (30°/s)}\\
\cline{2-9}
\hline
\vspace{0.5em}Rendering Type\vspace{0.5em} &Foveated&Full-view&Foveated&Full-view&Foveated&Full-view&Foveated&Full-view\\
\hline
\vspace{0.5em}Uncompressed 1:1\vspace{0.5em} &\cellcolor[HTML]{FFC0CB}544.19 Gbps&\cellcolor[HTML]{FC89AC}1.81 Tbps&	\cellcolor[HTML]{FFC0CB}906.99 Gbps&\cellcolor[HTML]{FC89AC}3.02 Tbps&\cellcolor[HTML]{FC89AC}2.26 Tbps&\cellcolor[HTML]{FC89AC}7.55 Tbps&\cellcolor[HTML]{FC89AC}8.16 Tbps&\cellcolor[HTML]{FC89AC}27.21 Tbps\\
\hline
Ultra Low-latency HEVC compressed 3:1&\cellcolor[HTML]{FFC0CB}181.39 Gbps&\cellcolor[HTML]{FFC0CB}604.66 Gbps&\cellcolor[HTML]{FFC0CB}302.33 Gbps&\cellcolor[HTML]{FC89AC}1 Tbps&\cellcolor[HTML]{FFC0CB}755.82 Gbps&\cellcolor[HTML]{FC89AC}2.51 Tbps&\cellcolor[HTML]{FC89AC}2.72 Tbps&\cellcolor[HTML]{FC89AC}9.06 Tbps\\
\hline
Low-latency compression 20:1&\cellcolor[HTML]{FFE4E1}27.20 Gbps&\cellcolor[HTML]{FFE4E1}90.69 Gbps&\cellcolor[HTML]{FFE4E1}45.34 Gbps&\cellcolor[HTML]{FFC0CB}151.16 Gbps&\cellcolor[HTML]{FFC0CB}113.37 Gbps&\cellcolor[HTML]{FFC0CB}377.91 Gbps&\cellcolor[HTML]{FFC0CB}408.14 Gbps&\cellcolor[HTML]{FC89AC}1.36 Tbps\\
\hline
Lossy compression 300:1&\cellcolor[HTML]{FCF3F1} 1.81 Gbps&\cellcolor[HTML]{FCF3F1} 6.04 Gbps& \cellcolor[HTML]{FCF3F1}3.02 Gbps&\cellcolor[HTML]{FFE4E1}10.07 Gbps& \cellcolor[HTML]{FCF3F1}7.55 Gbps&\cellcolor[HTML]{FFE4E1}25.19 Gbps&\cellcolor[HTML]{FFE4E1}27.20 Gbps&\cellcolor[HTML]{FFE4E1}90.69 Gbps\\
\hline
\end{tabular}
\label{TableI}
\begin{tablenotes}
   \item[] Data rate= 2 $\times$ FOV$\times$ BPP$\times$ APD$^2\times$  RR$\times$ \{1: for full view rendering, 0.3: foveated rendering\}/CR.
\end{tablenotes}
 \end{threeparttable}
\end{table*}

\subsection{Metaverse Requirements}

 \subsubsection{Latency}
The Metaverse requires a low end-to-end latency for an immersive experience. This is because latency can cause abrupt and discontinuous lag spikes in the display of pixels, causing motion sickness to the user, thus, disrupting the immersive experience. The total end-to-end latency for the Metaverse ranges from 7 to 20 ms \cite{cuervo2018creating,saad2019vision,bastug2017toward,bui2023game}, which is an aggregate of the following latencies: sensors, display, rendering computation, communication (a round-trip between the VR headset and the edge server), compression and decompression, and computing.
Real-time applications impose stringent computational latency requirements. However, the VR headset does not have the computational capabilities to render the Metaverse scene.
Consequently, the heavy computations are performed by an edge server nearby the user that performs low-latency computing.

 \subsubsection{Sensing and Localization}
For an immersive Metaverse experience, high-level accuracy localization and sensing are required \cite{tang2022roadmap}, such as positions of humans or objects, tracking of human eyes, etc. This will enable a real-time and accurate digital replica of the physical world in the Metaverse. 
In addition, the digitally replicated entities (also known as digital twins) need to be shared seamlessly and in real time with the users engaged in the Metaverse.
It is expected that the terahertz (THz) and visible light spectrum will be used for both communication and sensing \cite{xu2022full}, and therefore, meticulous attention to design integrated sensing and communication (ISAC) systems \cite{HVR2}.
In this paper, THz frequencies include sub-THz W-band (92-115 GHz), sub-THz D-band (130-175 GHz), and frequencies ranging from 175 GHZ to 10 THz.

 \subsubsection{Capacity}

To replicate a lifelike experience in the Metaverse, unprecedented high data rates are required to transmit the content \cite{xu2022full}. Furthermore, ultra-high capacity data rates are necessary to digitally emulate human visual perception.
Although tethered VR headset solutions allow the transmission of ultra-high data rates, it comes with a price of restricted mobility, lessened immersion, and tripping hazards \cite{chakareski2022toward}. Therefore, untether solution via wireless communication is the solution to fully experience the Metaverse.

\subsection{Wireless Data Rate Desiderata}
The Metaverse data rate requirements are determined by the following parameters \cite{cuervo2018creating}.

\subsubsection{Field Of View (FOV)} To emulate lifelike human vision, the VR headset should have the same FOV as the human eyes, that is, 210° horizontally and 135° vertically \cite{cuervo2018creating} \cite{andersen2002history}. Since there is an overlap of 114° between the two eyes, each eye has a view of 162°.
\subsubsection{Bits Per Pixel (BPP)} Human eyes can discern about 10 million colors with a contrast range of up to 1:$10^9$ \cite{cuervo2018creating} \cite{boitard2015evaluation}. To that end, VR headset must have a high dynamic range color using 96~bits per pixel.
\subsubsection{Angular Pixel Density (APD)} A normal vision acuity (20/20 vision) allows the eye to distinguish two contours separated by 1.75~mm at a distance of 6~meters (20~feet), that is, 60~pixels per degree \cite{cuervo2018creating,visual1984visual}.
    Recently, Meta stated that their VR headset Quest 2 has 21~pixels per degree, which is about 3 times lower than the basic 20/20 vision standard. Therefore, current VR headsets require further advancement to mimic lifelike displays.
\subsubsection{Refresh Rate (RR)} The refresh rate is the number of times the image refreshes on the screen per second. The human eye can track continuous movements at 30 degrees/second \cite{britannica1987sensory}. To avoid judder and motion sickness, the frame rate of the VR headset should be fast enough to avoid any abrupt discontinuous jumps in pixels related to a moving object.
Given a VR headset with an APD of 60 pixels per degree, the pixels corresponding to this image will move by 1800 pixels per second. Therefore, a frame rate of about 1800 Hz would guarantee
that there are no discontinuous pixel jumps. Recently, the company Asus announced their new gaming monitor \textit{The Asus ROG Swift 500 Hz}, which is the first monitor to reach a 500 Hz refresh rate \cite{asu}. We note that there are no comprehensive studies on the impact of refresh rate on the Metaverse experience.

\subsubsection{Foveal and Peripheral Vision} The foveal vision, which is located around the fovea (center of the eye), has a high visual acuity and is capable of discerning finer details in the scene, while the peripheral vision is far from the fovea and cannot discern fine details.
A VR headset with eye tracking can render a scene with full resolutions in the foveal and lesser resolutions in peripheral vision.
For instance, a 210° × 135° FOV, foveated rendering reduces the number of pixels by 70\% without  (significantly) affecting the user's perception \cite{patney2016towards}.

\subsubsection{Compression Rate (CR)} 
The CR represents a quantification of the relative reduction in the size of data produced by a data compression algorithm.
Lossless compression allows the original data to be perfectly reconstructed from the compressed data such as high-efficiency video coding (HEVC) with a CR of 2:1 to 3:1 \cite{sullivan2012overview}.
In contrast, lossy compression reduces the required data rates for transmitting frames while it might provide, according to the authors, an acceptable Metaverse experience to the user.
However, this comes with a price of an added delay proportional to the compressed data volume. Therefore, the compression rate has to be adjusted to meet the latency constraints of the VR headset.
We note that the effect of compression rates on the Metaverse experience has not been fully investigated.

%\subsubsection{Partial Computation} 
%The VR headset must have the pixel density and the refresh rate that is required for a lifelike experience,
%Therefore, the edge server that renders the frames must match the refresh rate of the VR headset.
%Assuming that the edge server GPU can render at the equivalent of 140 TFlops, then the VR headset will operate at a 180 Hz refresh rate
%which is less than the current 500 Hz refresh rate of The Asus ROG Swift gaming monitor. Consequently, the edge server GPU will compute at a refresh rate of 180 Hz and transmit only the rendered frames, while the remaining frames will be generated in the VR headset GPU using computational approximation image-based rendering (IBR). In that case, the transmission data rates will decrease.

Table~\ref{TableI} shows the data rates required to deliver a Metaverse experience according to different refresh rates, compressing rates, and rendering types. 
Note that since there are no comprehensive studies investigating the impact of refresh rate on the Metaverse experience, we cautiously use 1800 Hz as a maximum for refresh rates.

\subsection{Enabling Communication Technologies}
%As shown in Table~\ref{TableI}, the current wireless technologies cannot withstand the ultra-high data rates requirements of the Metaverse. 
5G is excepted to achieve a peak data rate of 20~Gbps \cite{tataria20216g,njoku2022role}.
On the other hand, Wi-Fi~7 is excepted to offer a maximum data rate of 5.8~Gbps, which is 2.4 times faster than the 2.4~Gbps achieved with Wi-Fi~6E \cite{intel}.
These wireless technologies cannot withstand the ultra-high data rates requirements of the Metaverse as shown in Table~\ref{TableI}.
Therefore, it is necessary to harness frequencies beyond the mmWave spectrum (30-90~GHz) \cite{chaccour2022can,wang2022meta,HVR}. 
Due to their large available bandwidth, optical and THz frequencies are regarded as key communication enablers to deliver the unprecedentedly high data rates required by the Metaverse \cite{lou2023coverage}. 
In addition to the ultra-high data rates provided by the THz frequencies, they require narrow pencil beamforming, thus, reducing the interference drastically.
%However, THz frequencies face several challenges. Such challenges include high path loss transmission, blockage-prone communication, and an accurate beam direction.
%However, THz frequencies are subject to high atmospheric transmission loss. 
In indoor Metaverse applications, although short distances enable high-rate transmission at THz frequencies, the Metaverse users’ bodies may lead to dynamic blockages over the THz links. This will negatively affect the immersive Metaverse experience. 
Consequently, when deploying only a THz-based network for Metaverse applications leads to several challenges, including user positioning, link blockage reduction, and reliability \cite{wang2022meta}.

Optical communications offer a huge opportunity to bring Metaverse applications to the general public owing to the advances in emerging optical communication technologies and the large unlicensed optical spectrum.
Optical fibers are the cornerstone of telecommunication infrastructure, and have been expanded to the backhaul and fronthaul components of the network. 
%Boosting the capacity of optical fibers has been a topic of interest by telecom providers. 
The usage of OWC is also maturing for backhaul/fronthaul and access applications. OWC can provide higher-speed connectivity in an underwater environment compared to any other technology. In the following, we show the various formats of fiber communication then OWC that can be used to satisfy the Metaverse capacity demand.

\begin{figure*}[ht!]
   \centering
    \includegraphics[width=0.99\linewidth] {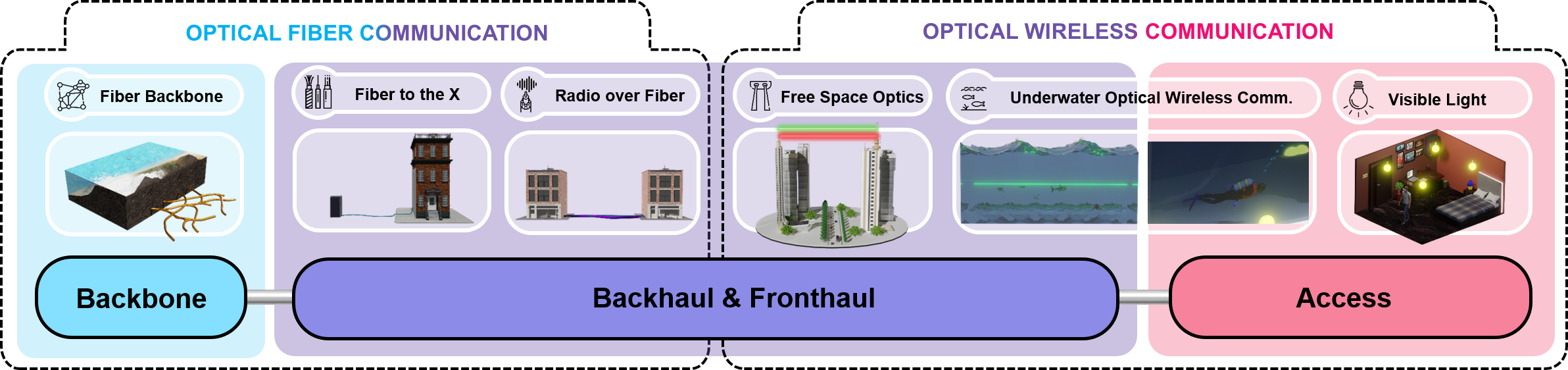}
    \caption {Optical fiber and OWC, including the backbone, backhaul/fronthaul, and access components, and the type of technologies of each component.}
    \label{opticalcomponents}
\end{figure*}

\section{The role of Optical Fiber Communication}\label{section3}
Here, we discuss the key advances that can be used to boost the capacity of fiber networks to support the Metaverse requirements.
\subsection{The Fiber Backbone}
Breakthroughs in optical fiber technologies have allowed continuously higher transmission rates and enabled various new applications, simultaneously boosting demand for higher capacity. Progress in techniques incorporating the use of different degrees of freedom of a lightwave, namely wavelength division multiplexing (WDM),  polarization division multiplexing (PDM) as well as the use of advanced optical modulation formats over single mode fibers (SMFs), has remarkably led to high bit-rate communications \cite{WinzerOPEX18}. Superchannels using parallel spatial paths has been proposed to continue the capacity growth in telecommunication systems and push toward
the limits \cite{WinzerOPEX18}. The concept is better known as space division multiplexing (SDM) and aims to use multicore fibers (MCF) or few-mode fibers \cite{RichardsonNatPhot13}. An MCF can be viewed as a superposition of multiple SMFs sharing the same cladding. At the same time, SDM over FMFs consists of encoding independent data streams over different light modes, known as linearly polarized (LP) modes. The latter concept is commonly known as mode division multiplexing (MDM). Both SDM over MCF and MDM are additions to existing multiplexing technologies, meaning that the same core can carry independent signals over multiple wavelengths and two orthogonal polarization states. The same mode can also be re-used at different wavelengths and polarization. Combining the two SDM approaches is possible by using a few-mode MCF. In such a case, the overall fiber capacity is scaled by the number of cores and the number of modes supported by each core. Specialty optical fibers \cite{RuschCommag18} carrying orbital angular momentum (OAM) are also promising for MDMs. A particular advantage of OAMs compared to LP modes is the ability to propagate over optical fibers with minimum intermodal crosstalk and random exchange of energy between modes, therefore not requiring digital signal processing (DSP) at the fiber output. OAM over fiber can enable direct fiber/optical wireless links. Unprecedented data rates beyond 10 Pbps (10.000 Tbps) have been reported thanks to SDM \cite{RademacherOFC20}. These data rates provide optical backbone infrastructures that can sustain the bandwidth demand of the emerging Metaverse era.

\subsection{Fiber to the X}

Fiber to the x (FTTX) is a last-mile solution, often referred to as fiber to the home.
Fiber access to the home and indoor environments is becoming widespread through passive optical networks (PONs) consisting of shared fiber systems installed by telecom service providers aiming to bring broadband network access to end users. Data rates provided by a single fiber in FTTX can reach 10 Gbps and beyond \cite{pagare2021design}. There is now emerging activity on fiber-to-the-room technologies so that fiber will ultimately be available in individual indoor spaces, especially for the Metaverse applications where high data capacity is required. Such links can provide orders of magnitude higher capacity compared to radio and copper-based wired links.
\subsection{Carrying Radio over Fibers}
Another short-range fiber solution that cater to the Metaverse data rates requirements is the radio over fiber (RoF), or RF over fiber, referring to a technology whereby light is modulated by an RF signal and transmitted over an optical fiber. RoF offers resilience to noise and electromagnetic interference, particularly to microwave signals also constrained by the free space transmission distances. Up to 10.5 Tbps data rate can be reached through RoF over a 20~km-long fiber \cite{zhu20221lambda}. Recently THz waves are also carried on fibers, similarly to microwave bands giving more hope to THz signals to travel over long distances without being affected by free space loss. The major limiting factors for RoF are chromatic dispersion, relatively low radio optical efficiency, and bandwidth inefficiency when using a small fraction of the available optical spectrum. Such challenges are being addressed in the literature (\cite{LimJLT21}, and references therein).
RoF links can provide high-throughput connectivity leveraging millimeter waves from access to the backhaul network.

\section{The Role of Optical Wireless Communication}\label{section4}
OWC is possible in different bands of the optical spectrum, namely the infrared (IR), visible, and ultraviolet (UV). Here, we mainly focus on OWC in the IR band, namely free space optics (FSO) and OWC in the visible range known as visible light communication (VLC).
\subsection{Free Space Optics}
FSO can deliver high-data-rate links \cite{TrichiliJOSAB20} and provide last-meter and last-mile fiber solutions without a physical fiber when fiber installation is not possible or costly. It can provide beyond Tbps connectivity in distances of hundreds of meters \cite{CiaramellaJSAC09}.
FSO can be quickly deployed and also can be dissembled and moved to other locations; therefore, it can provide high-speed connectivity for temporary Metaverse events. 
Since FSO is sensitive to atmospheric turbulence and weather conditions, it can be installed on top of RF links which can be used when the IR optical wireless link is unavailable.
\newline
By using spatially structured light, such as beams carrying OAM, FSO technology can be promising. 
Coping with these challenges and implementing high-speed FSO links can help bring the required capacity for the Metaverse without needing permanent fiber backhaul infrastructure.
Steerable FSO can be used for backhaul networks \cite{curran2017fsonet} where the steering of the beam is pre-configured and controlled between a set of transceivers. 
However, in Metaverse applications where the user’s VR headset transceiver can be in motion, maintaining a link becomes challenging.
Microsoft patented an
FSO-enabled VR headset \cite{cuervo2019mixed} with an in-band feedback mechanism, and laser beam motion tracking of the user’s transceiver VR headset \cite{rahman2018fso}.
However, it comes with several challenges, making FSO best suited for fronthaul/backhaul rather than access, especially for VR applications.

\begin{table*}[ht!]
\begin{threeparttable}[t]
\centering
\caption{\label{tab:Advances} Summary of Optical Communication Advances, Records, and challenges.}
\begin{tabular}{|m{32pt}|m{42pt}|m{66pt}|m{66pt}|m{110pt}|m{120pt}|}
\hline
Segment&Technology&Max Date Rates*&Max Distance**&Challenges&Enabling setting\\
\hline
Backbone&Fiber\newline Backbone&10.66 Pbps (13 km) \cite{RademacherOFC20}&17,107 km (51.5 Tbps) \cite{CaiOFC18}&- Nonlinearity\newline - Intermodal and intercore crosstalk for SDM \newline - Chromatic dispersion&-Advances in WDM and superchannels\newline- Combining SDM approaches through few mode MCFs\newline- Nonlinearity compensation \newline -Dispersion compensation techniques\\
\hline 
\multirow{4}{*}{\parbox{1cm}{Backhaul/\\Fraunthaul}}&FTTX&10 Gbps \cite{pagare2021design}&$\sim$100 m \cite{pagare2021design}&- Cost and maintenance&- WDM technology\\\cline{2-6}
&RoF& 10.5~Tbps (2 km) \cite{zhu20221lambda}  &100 km (5 Gbps) \cite{ShiuPJ20}&- Nonlinear distortion\newline- Chromatic dispersion &- Dual-polarization transmission\newline- Employing noise shaping technique\\\cline{2-6}
&FSO&1.28 Tbps (212 m) \cite{CiaramellaJSAC09}&20 km (20 Gbps) \cite{Taara}&- Sensitivity to turbulence\newline- Pointing errors&- WDM technology\newline - Incorporating adaptive tracking\\\cline{2-6}
&UWOC&3 Gbps (100.6 m) \cite{FeiOPEX22}&150 m (15 Mbps) \cite{BlueComm200}&- Underwater turbulence\newline - Pointing errors&- Using highly-sensitive detectors such as PMTs.\\\cline{2-6}
\hline
\multirow{5}{*}{Access}&LED VLC\newline&15.73 Gbps (1.6 m) \cite{BianJLT19}&100 m (1 Gbps) \cite{WangOE16}&- Sensitivity to ambient light&- Using $\mathrm{\mu}$-LED arrays\newline- Using advanced modulation formats.\\\cline{2-6}
&LD VLC&40.665 Gbps (2 m) \cite{WeiOPEX19}&100 m (6 Gbps) \cite{QinACPC21}&- Sensitive to user mobility \newline - Safety issues for high power levels depending on the visible band &- Multiplexing LDs with different colors \newline -Polarization multiplexing\\\cline{2-6}
&SLD VLC&4.57 Gbps \cite{LiCrystals22}&1 m \cite{LiCrystals22}&- Device sensitivity&- Using GaN-based SLD\\\cline{2-6}
&LED UWOC&6.9 Gbps (2 m) \cite{ChengOPEX22}&46 m ($\sim$ Mbps)  \cite{ShenOptCom19}&- Sensitivity to propagation effects&- Using mini-LED\newline - Combining multiple colors\\\cline{2-6}
&LD UWOC&30 Gbps (2.5 m) \cite{TsaiSREP19} &50 m (80 Mbps) \cite{ZhouJLT22}&- Fulfilling the PAT requirements\newline - Sensitivity to turbulence&- Extending the FOV of PDs\newline- Active tracking\\\cline{2-6}
\hline
\end{tabular}
  \begin{tablenotes}
 \item[*]  with associated distance for the maximum demonstrated data rate. ** with associated data rate for the maximum reported distance.
  \end{tablenotes}
  \end{threeparttable}
\end{table*}

%Besides the various terrestrial applications, IR optical wireless links are already being used for inter-satellite crosslinks in broadband space constellations and are also excepted to argument existing microwave satellite feeder links in the near future \cite{HemmatiProcIEEE11}. Since laser beams can be subject to propagation effects at the lower layers of the atmosphere, site diversity can ensure continuous operations if the transmitted signals are blocked in one or multiple locations since weather conditions are different in various locations.\newline 
%In addition to outdoor and space applications, IR FSO-like links can be used to connect racks of servers indoors, which can significantly reduce the cabling complexity and increase the overall channel capacity. \textbf{Metaverse ?}
\subsection{Visible Light Communication}
Indoor OWC, also known as Light Fidelity (LiFi) or VLC when visible light
frequency (400-800 THz) are used, has seen tremendous progress over the last decade \cite{HaasPTRSA20}. 
VLC, co-existing with lightning, can offer multi-Gbps wireless connectivity and augment the capacity provided by radio RF-based WiFi systems \cite{janjua2015going,lee20152}. 
A combination of RF systems, which can provide reliable coverage and a moderate capacity (as detailed in Section \ref{Configuration}), and OWC systems that provide ultra-high capacity and much-needed additional wireless spectrum, is one of the key enabling technologies to implement future Metaverse applications. 
Recent reports have shown that light emitting diode (LED)-based VLC can provide data rates beyond 15 Gbps over a meter scale range when incorporating high-order modulation formats such as M-QAM orthogonal frequency division multiplexing (OFDM). 

Using laser diodes (LDs) is also possible for Metaverse indoor applications requiring low user mobility and can provide larger throughputs than LED-based solutions. The recent introduction of super-luminescent diodes (SLDs) in VLC systems can provide high power signals at low coherence \cite{shen2019group,alatawi2018high}.\newline
Using light signals to communicate indoors also offers unique security features. Contrary to the widely used WiFi, VLC signals cannot penetrate walls, making such technology secure against neighboring-room interceptors. LiFi can be further used for Metaverse applications in RF-sensitive environments, such as healthcare departments. \newline
VLC requires line-of-sight (LOS) alignment between transmitter and receiver beams, which becomes challenging when the Metaverse users are in motion. 
Therefore, the use of multiple optical small cells placed in different locations in the Metaverse room can provide high data rate coverage to multiple Metaverse users even when they are in motion. Also, NLOS issues can be alleviated by using reflecting metasurfaces in room with wavefront shaping (see Section~\ref{section6}). 
Another promising solution is a VR headset especially dedicated to VLC applications. 
It is able to maintain connectivity and cope with the VR user’s random head movement \cite{khan2019visible} (see Section~\ref{section5}).
%It ought to be mentioned that depending on the considered visible band, safety restrictions should be considered when using LD for high power levels. 
Finally, VLC can potentially be used for positioning and localization in ream-time for Metaverse applications. 
%, particularly in indoor environments. %The satellite positioning systems cannot meet the requirements of accurate indoor environments %\cite{PhamOPEX19}. 
%Visible light positioning can co-exist with communication and provide accurate indoor localization \cite{PhamOPEX19}. Visible light can also enable indoor navigation that can be required for real-time Metaverse applications.

%Communication optical fibers can also be used simultaneously for distributed acoustic sensing (DAS) and distributed temperate sensing (DTS) purposes while carrying information signals, as demonstrated in the literature \cite{GoogleSensing22}. Incorporating sensing functions can benefit the Metaverse applications. In some cases, it is possible to accomplish sensing functions without the need for a dedicated sensing unit, and instead, the communication signals can serve as tools to observe the changes through an interferometric approach \cite{IpJLT22}\newline

\subsection{Underwater Wireless Optical Communication}
UWOC, over visible wavelength, has seen considerable attention as it offers high data rates over tens of meters. Compared to acoustic communication suffering from long delays due to the low sound speed in the water, UWOC can be seen as a low-latency and stealth technology that limits the noise pollution that might affect underwater marine life.\newline
Up to 30 Gbps data rate is now possible using LDs over a few meters distance. Lower data rates are reached when LEDs are used instead. But the achievable rates are increasing thanks to color multiplexing and various digital signal processing (DSP) techniques. UWOC is constrained by various challenges imposed by the statistically-random properties of harsh underwater environments. Many solutions have been proposed to cope with these issues. Solutions such as diffused light sources and extended FOV photonic detectors can improve the performance of UWOC links.
%In an underwater environment, since satellite positioning signals are unavailable, optical signals can be used to provide the locations of submerged devices. UWOC signals can be exploited equally to communicate with the internet of underwater (IoUT) devices that can be part of the Metaverse experience. In addition, several Metaverse applications, such as underwater virtual tourism and explorations, take place in an aquatic environment. UWOC will enable such applications with the reacquired data rates. \newline %IoUT devices can also be self-powered and harvest energy from UWOC signals.\newline 

Underwater Metaverse has great potential with several promising applications such as underwater tourism, training, recreation swimming, snorkeling, cybersickness, and underwater agriculture \cite{tang2022roadmap}.
In a Metaverse-based swimming pool, users can explore the underwater world \cite{swimvr}; for training, astronauts used it to emulate conditions experienced during their mission \cite{sinnott2019underwater}, and
for recreational swimming, swimmers can access in real-time their data and stats regarding their swimming abilities such as record lines, speed, and swimming forms \cite{yamashita2016aquacave}.
It can also allow snorkeling with a lifelike experience \cite{snorkel}.
Another aspect of underwater Metaverse is curing cybersickness for people who suffer from it and cannot handle dry ground Metaverse experience \cite{fauville2021effect}. 
Recently, a company named Nemo’s Garden used the underwater Metaverse for underwater farming \cite{NG}.
The company requested Siemens to create an accurate
digital twin of its greenhouse domes in the Metaverse, before starting the
real-life construction. Therefore, the divers were able to practice in the Metaverse. It also allowed them to foresee any possible challenges and address them.
By modeling the greenhouse environment in the Metaverse,
the company was able to develop, adapt, and control its underwater biospheres at scale \cite{Siemens}.

\begin{figure}[t]
    \centering
    \includegraphics[width=0.95\linewidth] {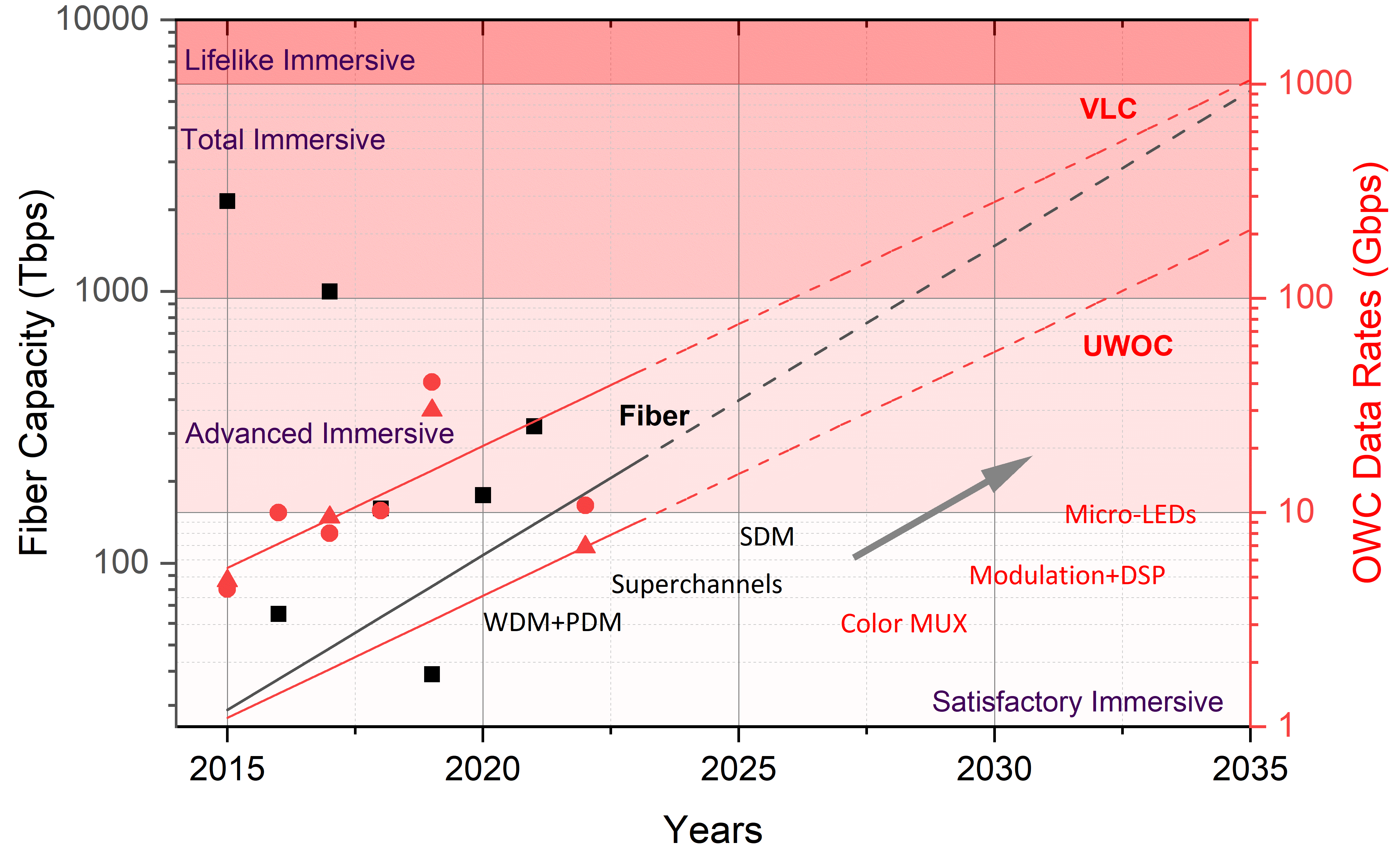}
    \caption {Anticipated data rates evolution of the various optical communication forms (fiber backbone, VLC, and UWOC). Squares, circles, and triangles correspond to data rates achieved in research testbeds over fiber \cite{PuttnamECOC15,KobayashiOFC17,RademacherOFC18, CorcoranNatCom20, GaldinoPTL20,PuttnamOFC21}, VLC \cite{WeiOPEX19, RetamalOPEX15, BianECOC18, ChunJLT16, GutemaJLT22}, and UWOC links \cite{ChengOPEX22,TsaiSREP19,OubeiOPEX15, KongOPEX17}}, respectively. The lines correspond to a 30\% technology evolution approximated  based on the evolution of the data rate of optical communication at a rate of 10,000 times in 30 years \cite{Liu2019}.
    \label{fig:DR}
\end{figure}

An illustration of all the aforementioned components and technologies is shown in Fig.~\ref{opticalcomponents}. A summary of the maximum reported data rates and propagation distances based on notable optical communication demonstrations or commercially available products is given in Table~\ref{tab:Advances}. We note that we intentionally excluded values from studies conducted in perfectly controlled underwater environments, such as those using mirrors to extend the propagation path over a water tank used to mimic an underwater channel. Challenges and perspectives for each of the technologies are briefly mentioned.
Finally, a prediction of anticipated data rates evolution of fiber backbone, VLC, and UWOC are depicted in Fig.\ref{fig:DR}

%=========================================================================
%=========================================================================
%=========================================================================
\begin{figure*}[h]
    \centering
    \includegraphics[width=0.99\linewidth] {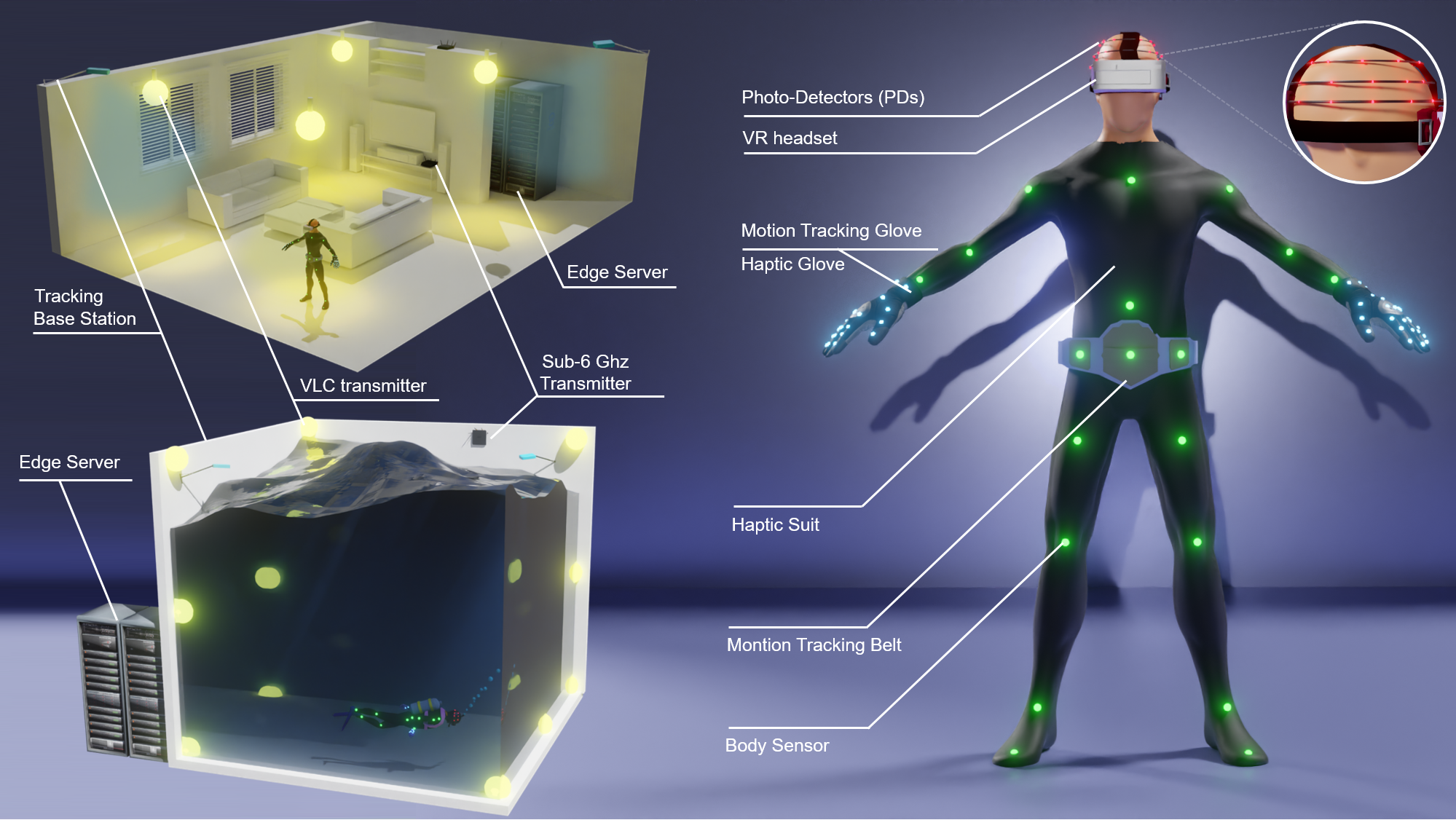}
    \caption {Illustration of the Metaverse arena (MetaChamber and MetAquarium), connectivity, and the necessary equipment for the Metaverse.}
  \label{Metaverse}
\end{figure*}

\section{Metaverse Configuration, Arena, and Equipment}\label{Configuration}\label{section5}
In the section, we detail the Metaverse configuration, arenas, connectivity, and equipment as illustrated in Fig.~\ref{Metaverse}.
\subsection{Configuration}
We will illustrate two Metaverse configurations: an indoor and an underwater configuration. We will use throughout the rest of the paper the term \textit{Metaverse arena} referring to the location in which the Metaverse experience takes place.
In the indoor configuration, the Metaverse arena is termed \textit{MetaChamber}, in which the user explores the virtual universe in a dedicated or regular room. 
In the underwater configuration, the Metaverse arena is termed \textit{MetAquarium} in which the user explores the virtual universe in an aquatic environment, which can be a pool or a dedicated aquatic container.

\subsection{Metaverse Arena}
The Metaverse arena is equipped with LED-, LD-, or SLD-based VLC transmitters mounted on the walls and ceiling. VLC transmitters will create several optical attocells in the Metaverse arena. 
Optical attocells have smaller cell sizes than femtocells, but they have full functionality offered by cellular networks such as handover, multiple access, and full duplexing \cite{chen2015downlink}.
Consequently, attocell footprints will provide coverage primarily for connectivity and secondary for illumination. Users are connected to their closest VLC transmitter; therefore, they might be connected to another VLC transmitter when they move or swim across the Metaverse arena. 
In addition, sub-6~GHz and mmWave access points will be mounted on the Metaverse arena walls and ceiling.
Users' spatial locations are tracked by tracking base stations, and they will be placed on the four upper corners of the Metaverse arena. Although two tracking base stations are enough to track the user's movements, the other two will increase the accuracy of the tracking.
These tracking base stations are commercially available with a tracking accuracy of less than 1 mm error \cite{kuhlmann2023static,Streamvr,valve}.

\subsection{Dual Connectivity}
The VR headset will enable dual connectivity for the downlink (from the edge server to the user) and single connectivity for the uplink (from the user to the edge server).
For downlink communications, dual connectivity is used to stream parallelly two Metaverse content layers: an immersive layer and a backup layer. 
The immersive layer contains raw uncompressed content data with full view that are transmitted via VLC technology since its data rate requirements are extremely high.
However, VLC communication might be subject to an outage due to blockage and path loss. Therefore, the backup layer will deliver compressed content data to ensure seamless connectivity. The compression rate will depend on the achievable data rate. In this layer, data are transmitted in sub-6~GHz frequencies and, when possible, mmWave.
For uplink communications, only single connectivity using sub-6~GHz is required with low data rates since it is used to transmit to the edge server small-size data such as control information, headset orientation, movements, the position of the user, etc.
\subsection{Metaverse Equipment}
\subsubsection{VR Headset}
VR headset is a head-mounted device that provides the rendered Metaverse scene to the user.
VR headset can determine in real-time the user’s head orientation and position. This feedback tracking information is transmitted using the uplink and then used by the edge server to adapt the scene rendering accordingly.
The head orientations and position feedback tracking information are also used to maintain the optical wireless link in a direct line-of-sight (LOS) between the transmitter and the VR headset user. To this end, optical transmitters are placed on a steerable platform and directed toward the user’s VR headset using the tracking information.
On the other hand, a hemispherical cap can be added to the upper portion of the VR headset covered with multiple photodetectors (PDs) on its surface \cite{chakareski2022toward}. These PDs can receive and decode, in parallel, the optical signal from the LED/LD/SDL on their surfaces as shown in Fig.~\ref{Metaverse}. In addition, the decoded incident signals at each PD are combined using diversity combining techniques to maximize the performance. 
\subsubsection{Tracking Motion Equipment}
Tracking sensors are placed on the main joints of the body, as illustrated in Fig.~\ref{Metaverse}.
The sensors track gravitational pull, rotation, and position to capture movements fully. In addition, tracking sensors are placed around a belt to capture the rotation of the hips. 
Finally, the glove will have sensors to record the movement of the hands and fingers. All the tracking motion information is then transmitted to the edge server to capture the motion in real-time and dynamically render the scene.

\subsubsection{Haptic Equipment}
For complete immersion, users can wear haptic gloves and a suit. Haptic gloves use actuators and motors to simulate the resistance, feeling, and vibration as tactile feedback. For instance, this will allow users to feel the object when grabbing or touching it in the Metaverse world. Similarly, haptic suits provide haptic feedback to the user’s body.

\subsubsection{Additional Equipment}
In addition to the aforementioned equipment, here is a non-exhaustive list of equipment to enhance the Metaverse experience: a Metaverse treadmill that allows users to perform walking or running motions in any direction, 
 an emotion recognition device to detect the emotion expressed by the user, and a
oxygen cylinder for the MetAquarium.
\begin{itemize}
\item Metaverse treadmill: It allows users to perform walking or running motion in any direction with a 3DoF. Therefore, allowing them to move the limiting boundaries of the Metaverse room. Note that some treadmills require special shoes.
\item Emotion recognition apparatus: This will detect the emotion expressed by the user and translate it into its avatar in the Metaverse. This can be carried out via emotion recognition sensors or via deep Learning training for facial emotion detection from images or videos taken by cameras \textit{in situ}.
\item Oxygen cylinder: If the user will dive into the MetAquarium and not stay on the surface, it is mandatory to have an Oxygen cylinder to experience an immersion experience. The duration of the experience will depend on the oxygen cylinder sizes and capacities.
\item Waterproof equipment: All the underwater equipment should be waterproof such as the VR headset, sensors, actuator, detector, camera, communication devices, etc.
\end{itemize}

\subsection{Edge Computing Server}
The edge server is installed \textit{in situ} close to the Metaverse arena, and they are connected together with an optical fiber link. The edge server has high-end storage and computing capabilities to lessen the computation burden on the VR headset and minimize communication latency. More importantly, the edge server will be responsible for rendering the scene according to the type of layer (immersive or backup), visualization (full view or foveated), degree of navigation (6 degrees of freedom [6DoF] or 3DoF), and processing (Real-time interactive or cache-based content).
\begin{itemize}
\item Layer: Depending on the technology used and data rates of the transmission, the edge server determines if the immersive layer or backup layer content should be transmitted.
\item Visualization: 
Depending also on the technology used and data rates of the transmission, edge server can either render a full view or a foveated view of the scene.
The edge server will collect tracking information from the VR headset and render the scene with lifelike details of the zone gazed at by the user and reducing the rendering quality in the peripheral vision.
\item Degree of navigation:
When delivering 3DoF content, the edge server must collect the user’s head positions and orientations (and possibly hand positions) to render the scene accordingly. However, when delivering 6DoF content, the edge server must also collect the user’s whole body movements, that is, head, hand, body, and joint movements.
\item Processing:
The type of processing will differ from interactive content, in which the user can interact in real-time with the Metaverse world and other users, or cache-based content, in which the user is an observer using the previously rendered data of the scene. In that case, the user’s action cannot alter or change the Metaverse scene. 
\end{itemize}

%== ==== ===== === === === === ===== ===== ===== ==== ======
%=========================================================================
%=========================================================================
%=========================================================================

%=========================================================================
%=========================================================================
%=========================================================================

\section{Key Enablers and Future Research Directions}\label{section6}

%\subsection{Supporting User Mobility with Ultraviolet Light}
%Tapping on the scattering of ultraviolet (UV) light, the \textcolor{red}{LOS} requirement of OWC can be tolerated. UV signals can reach the detector even if the main \textcolor{red}{LOS} component is obstructed, as demonstrated in several experiments reported in the literature, particularly in the C band (200-280 nm). The primary challenge of introducing such technology for human use would be exposure to harmful UV light. However, using UV light can be useful if a particular application requires wearing UV-protective haptic equipment. %\cite{VavoulasIEEECOMST19}
\subsection{Optical Reflecting Surfaces and Wavefront Shaping}
%The emergence of intelligent reflecting surfaces (IRSs) can improve the performance of FSO and LiFi. For instance, IRSs can ease the \textcolor{red}{LOS} requirements for LiFi by mitigating blockage and obstructions and therefore tolerate mobility, and random device orientations \cite{IRSLiFi22}. IRSs for OWC can be made either by metasurfaces or arrays of mirrors.
%Beyond communication, IRS can enable the integration of the sensing and localization features \cite{IRSLiFi22}.

When using VLC, dynamic blockages (caused by objects, by other users, or self-induced) prevent the LOS link to be established when users move in the Metaverse arena.
To circumvent this, non-line-of-sight (NLOS) links can be leveraged by using the light reflected off a material present in the room. However, the proposed solutions to alleviate this challenge compensate for the diffused losses by either increasing the system power or by avoiding the diffuse reflection.
The emergence of intelligent reflecting surfaces (IRS) can improve the performance of FSO and LiFi \cite{IRSLiFi22}. IRSs for OWC. This can be made either by metasurfaces or arrays of mirrors to make the propagation environment fully controllable with low cost and power.
In addition, wavefront shaping can be used to enhance the intensity of a diffuse NLOS link by allowing light to be focused through and inside opaque materials by controlling the wavefront of the light \cite{cao2019reconfigurable}. This will enhance the light intensity and maximize the scattered optical power at the receiver.
IRS and wavefront shaping can ease the LOS requirements for LiFi by mitigating blockage and obstructions in the Metaverse arena and therefore tolerate mobility, and random user and device orientations.
Beyond communication, IRS can enable the integration of the sensing and localization features \cite{IRSLiFi22}.

\subsection{Hollow Core Fiber Deployment}

Another promising solution is hollow core fiber (HCF) which can achieve light propagation at more than 99.8\% of the light speed in a vacuum.  
Compared to solid-core fibers, they allow approximately a 50\% increase in speed and a 33\% reduction in latency \cite{poletti2013towards}.
These speed improvements and latency savings are extremely significant in latency-sensitive Metaverse applications. 
In addition, HCF is backward compatible with the existing fiber solutions. It can be used in the backbone but also the backhaul and fronthaul.
Microsoft recently showed its interest in this technology and has acquired a HCF start-up, Lumenisity, to optimize its global cloud infrastructure further \cite{blogHCF}. However, this solution comes with several challenges that must be addressed for large-scale deployment.  %%%
%\subsection{Toward a Quantum Encrypted Internet}
%Several hundred-kilometer quantum key distribution (QKD) operations are now possible.
%FSO is also likely to play an essential role in the future of quantum information networks, as exemplified by steady progress in single-photon level QKD demonstrations.
%Combining QKD with FSO and fiber infrastructure can significantly improve security compared to classical encryption techniques that can be easily broken with future quantum computers. Satellite-based QKD is now possible and can contribute to the upcoming Quantum Internet paradigm. 
%\subsection{Wireless Energy Transfer}
%Power transfer can co-exist with optical communication. The concept is commonly known as simultaneous lightwave and information transfer (SLIPT) in OWC systems and power over fiber (PoF) in optical fibers. VLC signals, for example, can extend the battery life of some sensors equipped with solar panels to harvest power. Transferring power over fibers has been long demonstrated.
%\textbf{PAPER is too long, therefore we may need to remove this!}
%\subsection{Leveraging Machine Learning}
%There has been tremendous progress in using machine learning (ML) and related algorithms in optical wireless communication.
%ML algorithms can be practical for site-specific VLC channel modeling. 
%One other use case of ML is enabling switching between RF and FSO links in hybrid radio/optical systems. 
%Harnessing the power of ML can be beneficial for optical fiber communication by enabling modulation format identification and nonlinearity compensation.
\subsection{Hybrid THz/VLC Networks in the Access Segement}
Alongside the optical frequencies, THz is also considered a potential candidate to provide the extremely high data rate required by the Metaverse due to the large available bandwidth. In addition, THz frequencies have the potential for indoor positioning. 
Although THz frequencies are highly prone to blockage, which can negatively affect the immersive Metaverse experience, they can be used in a hybrid VLC-THz configuration to leverage both technologies for Metaverse applications \cite{wang2022meta}.
Depending on the configuration, THz can provide an alternative and accurate positioning service while VLC provides the higher data rates, and \textit{vice versa}. 
THz/VLC network can jointly leverage both technologies to provide the required high data rate for an immersive Metaverse experience and accurate positioning for Metaverse users.

\subsection{Analog Neuromorphic Optical Computing at the Edge}
The edge server in the Metaverse arena is required to perform high computation operations of the massive data with a low latency to render the Metaverse virtual scene. However, there is a fundamental bottleneck related to computing hardware in terms of speed and power consumption due to the explosive growth of data generated by the Metaverse.
Brain-inspired, also termed \textit{neuromorphic}, computing offers extremely high power efficiency and computational capabilities; but, it still relies on electronic components whose speed and energy are limited by RF crosstalk, Joule heating, and capacitance.
Optical neuromorphic computing overcomes this bottleneck by processing information using light and benefiting from the unique properties of photons such as broad bandwidth, low latency, and high energy efficiency \cite{wu2022analog}.
This will enable analog optical neuromorphic computing at the edge to process the massive data required to render a lifelike Metaverse world with much higher bandwidths and data rate as well as lower latency compared to their electronic counterpart.
Although current techniques of analog optical computing have shown the unique potential of light (such as speed, data parallelization, and low power consumption) \cite{de2019photonic}, it is still challenging to implement them due to the lack of suitable integrated architectures and integrated photonics devices.

\section{Conclusion}
In summary, we provided a holistic overview of the potential contributions of optical communication in the upcoming Metaverse era.
We first presented Metaverse requirements in terms of capacity and latency; then, introduced ultra-high data rates requirements for various Metaverse experiences.
We put forward the contributions of several fiber solutions needed in the backbone, backhaul, and access infrastructures; and showed how VLC will provide ultra-high data rates communications, accurate localization and sensing, and secure indoor coverage for the Metaverse applications.
We also demonstrated how the growing capacity of the various optical communication systems will continuously withstand the Metaverse requirements.
In addition, we proposed a comprehensive set of configurations, connectivity, and equipment necessary for an immersive Metaverse experience.
Finally, future exciting optical research directions are presented such as analog neuromorphic optical computing, hollow core fiber deployment, and optical reflecting surfaces.
\bibliographystyle{IEEEtran}
\bibliography{References.bib}

\begin{thebibliography}{10}
\providecommand{\url}[1]{#1}
\csname url@samestyle\endcsname
\providecommand{\newblock}{\relax}
\providecommand{\bibinfo}[2]{#2}
\providecommand{\BIBentrySTDinterwordspacing}{\spaceskip=0pt\relax}
\providecommand{\BIBentryALTinterwordstretchfactor}{4}
\providecommand{\BIBentryALTinterwordspacing}{\spaceskip=\fontdimen2\font plus
\BIBentryALTinterwordstretchfactor\fontdimen3\font minus
  \fontdimen4\font\relax}
\providecommand{\BIBforeignlanguage}[2]{{%
\expandafter\ifx\csname l@#1\endcsname\relax
\typeout{** WARNING: IEEEtran.bst: No hyphenation pattern has been}%
\typeout{** loaded for the language `#1'. Using the pattern for}%
\typeout{** the default language instead.}%
\else
\language=\csname l@#1\endcsname
\fi
#2}}
\providecommand{\BIBdecl}{\relax}
\BIBdecl

\bibitem{tang2022roadmap}
F.~Tang, X.~Chen, M.~Zhao, and N.~Kato, ``The roadmap of communication and
  networking in 6g for the metaverse,'' \emph{IEEE Wireless Communications},
  2022.

\bibitem{chang20226g}
L.~Chang, Z.~Zhang, P.~Li, S.~Xi, W.~Guo, Y.~Shen, Z.~Xiong, J.~Kang,
  D.~Niyato, X.~Qiao \emph{et~al.}, ``6g-enabled edge ai for metaverse:
  Challenges, methods, and future research directions,'' \emph{Journal of
  Communications and Information Networks}, vol.~7, no.~2, pp. 107--121, 2022.

\bibitem{bui2023game}
V.-P. Bui, S.~R. Pandey, F.~Chiariotti, and P.~Popovski, ``Game networking and
  its evolution towards supporting metaverse through the {6G} wireless
  systems,'' \emph{arXiv preprint arXiv:2302.01672}, 2023.

\bibitem{du2022exploring}
H.~Du, J.~Wang, D.~Niyato, J.~Kang, Z.~Xiong, X.~S. Shen, and D.~I. Kim,
  ``Exploring attention-aware network resource allocation for customized
  metaverse services,'' \emph{IEEE Network}, 2022.

\bibitem{GlobalData}
{GlobalData}, ``Metaverse market size, share, trends, analysis and forecasts
  2022-2030,''
  https://www.globaldata.com/store/report/metaverse-market-analysis/.

\bibitem{xu2022full}
M.~Xu, W.~C. Ng, W.~Y.~B. Lim, J.~Kang, Z.~Xiong, D.~Niyato, Q.~Yang, X.~S.
  Shen, and C.~Miao, ``A full dive into realizing the edge-enabled metaverse:
  Visions, enabling technologies, and challenges,'' \emph{IEEE Communications
  Surveys \& Tutorials}, 2022.

\bibitem{khan2019visible}
M.~Khan and J.~Chakareski, ``Visible light communication for next generation
  untethered virtual reality systems,'' in \emph{2019 IEEE International
  Conference on Communications Workshops (ICC Workshops)}.\hskip 1em plus 0.5em
  minus 0.4em\relax IEEE, 2019, pp. 1--6.

\bibitem{HVR}
{Huawei Technologies Co., Ltd.}, ``{Huawei} {iLab} {VR} technology white paper.
  {Cloud VR} bearer networks,''
  https://www-file.huawei.com/-/media/corporate/pdf/ilab/cloud vr oriented
  bearer network white paper en v2.pdf, 2017.

\bibitem{wang2022meta}
Y.~Wang, M.~Chen, Z.~Yang, W.~Saad, T.~Luo, S.~Cui, and H.~V. Poor,
  ``{Meta-reinforcement learning for reliable communication in THz/VLC wireless
  VR networks},'' \emph{IEEE Transactions on Wireless Communications}, vol.~21,
  no.~9, pp. 7778--7793, 2022.

\bibitem{cuervo2018creating}
E.~Cuervo, K.~Chintalapudi, and M.~Kotaru, ``Creating the perfect illusion:
  What will it take to create life-like virtual reality headsets?'' in
  \emph{Proceedings of the 19th International Workshop on Mobile Computing
  Systems \& Applications}, 2018, pp. 7--12.

\bibitem{saad2019vision}
W.~Saad, M.~Bennis, and M.~Chen, ``A vision of 6g wireless systems:
  Applications, trends, technologies, and open research problems,'' \emph{IEEE
  Network}, vol.~34, no.~3, pp. 134--142, 2019.

\bibitem{bastug2017toward}
E.~Bastug, M.~Bennis, M.~M{\'e}dard, and M.~Debbah, ``Toward interconnected
  virtual reality: Opportunities, challenges, and enablers,'' \emph{IEEE
  Communications Magazine}, vol.~55, no.~6, pp. 110--117, 2017.

\bibitem{KhanArxiv22}
L.~U. Khan, Z.~Han, D.~Niyato, E.~Hossain, and C.~S. Hong, ``Metaverse for
  wireless systems: Vision, enablers, architecture, and future directions,''
  \emph{ArXiv}, vol. abs/2207.00413, 2022.

\bibitem{wu2022analog}
J.~Wu, X.~Lin, Y.~Guo, J.~Liu, L.~Fang, S.~Jiao, and Q.~Dai, ``Analog optical
  computing for artificial intelligence,'' \emph{Engineering}, vol.~10, pp.
  133--145, 2022.

\bibitem{HVR2}
{Huawei Technologies Co., Ltd.}, ``The role of connectivity in building the
  metaverse,''
  https://www.huawei.com/us/huaweitech/publication/202207/ict-pave-way-metaverse,
  2022.

\bibitem{chakareski2022toward}
J.~Chakareski, M.~Khan, and M.~Yuksel, ``Toward enabling next-generation
  societal virtual reality applications for virtual human teleportation: A
  novel future system concept and computation-communication-signal
  representation trade-offs,'' \emph{IEEE Signal Processing Magazine}, vol.~39,
  no.~5, pp. 22--41, 2022.

\bibitem{andersen2002history}
S.~R. Andersen, ``The history of the ophthalmological society of copenhagen
  1900--50,'' \emph{Acta Ophthalmologica Scandinavica}, vol.~80, pp. 6--17,
  2002.

\bibitem{boitard2015evaluation}
R.~Boitard, R.~K. Mantiuk, and T.~Pouli, ``Evaluation of color encodings for
  high dynamic range pixels,'' in \emph{Human Vision and Electronic Imaging
  XX}, vol. 9394.\hskip 1em plus 0.5em minus 0.4em\relax SPIE, 2015, pp.
  532--540.

\bibitem{visual1984visual}
C.~A. C. O.~U. Visual, F.~Committee \emph{et~al.}, ``Visual acuity measurement
  standard,'' \emph{Ital J Ophthalmol}, vol.~2, pp. 5--19, 1984.

\bibitem{britannica1987sensory}
E.~Britannica, ``Sensory reception: human vision: structure and function of the
  human eye,'' \emph{Encyclopedia Brittanica}, vol.~27, p. 179, 1987.

\bibitem{asu}
{ASUS}, ``The rog swift 500hz shatters boundaries with its ultra-fast panel,''
  https://rog.asus.com/articles/news/the-rog-swift-500hz-shatters-boundaries-with-its-ultra-fast-panel/.

\bibitem{patney2016towards}
A.~Patney, M.~Salvi, J.~Kim, A.~Kaplanyan, C.~Wyman, N.~Benty, D.~Luebke, and
  A.~Lefohn, ``Towards foveated rendering for gaze-tracked virtual reality,''
  \emph{ACM Transactions on Graphics}, vol.~35, no.~6, pp. 1--12, 2016.

\bibitem{sullivan2012overview}
G.~J. Sullivan, J.-R. Ohm, W.-J. Han, and T.~Wiegand, ``Overview of the high
  efficiency video coding {(HEVC)} standard,'' \emph{IEEE Transactions on
  circuits and systems for video technology}, vol.~22, no.~12, pp. 1649--1668,
  2012.

\bibitem{tataria20216g}
H.~Tataria, M.~Shafi, A.~F. Molisch, M.~Dohler, H.~Sj{\"o}land, and
  F.~Tufvesson, ``{6G} wireless systems: Vision, requirements, challenges,
  insights, and opportunities,'' \emph{Proceedings of the IEEE}, vol. 109,
  no.~7, pp. 1166--1199, 2021.

\bibitem{njoku2022role}
J.~N. Njoku, C.~I. Nwakanma, and D.-S. Kim, ``The role of {5G} wireless
  communication system in the metaverse,'' in \emph{2022 27th Asia Pacific
  Conference on Communications (APCC)}.\hskip 1em plus 0.5em minus 0.4em\relax
  IEEE, 2022, pp. 290--294.

\bibitem{intel}
{Intel}, ``Wi-fi 7,''
  www.intel.com/content/www/us/en/products/docs/wireless/wi-fi-7.html.

\bibitem{chaccour2022can}
C.~Chaccour, M.~N. Soorki, W.~Saad, M.~Bennis, and P.~Popovski, ``Can
  {Terahertz} provide high-rate reliable low-latency communications for
  wireless {VR}?'' \emph{IEEE Internet of Things Journal}, vol.~9, no.~12, pp.
  9712--9729, 2022.

\bibitem{lou2023coverage}
Z.~Lou, B.~E.~Y. Belmekki, and M.-S. Alouini, ``Coverage analysis of hybrid
  {RF/THz} networks with best relay selection,'' \emph{IEEE Communications
  Letters}, 2023.

\bibitem{WinzerOPEX18}
P.~J. Winzer, D.~T. Neilson, and A.~R. Chraplyvy, ``Fiber-optic transmission
  and networking: the previous 20 and the next 20 years {[Invited]},''
  \emph{Optics Express}, vol.~26, no.~18, pp. 24\,190--24\,239, Sep 2018.

\bibitem{RichardsonNatPhot13}
D.~J. Richardson, J.~M. Fini, and L.~E. Nelson, ``Space-division multiplexing
  in optical fibres,'' \emph{Nature Photonics}, vol.~7, pp. 354--362, 2013.

\bibitem{RuschCommag18}
L.~A. Rusch, M.~Rad, K.~Allahverdyan, I.~Fazal, and E.~Bernier, ``Carrying data
  on the orbital angular momentum of light,'' \emph{IEEE Communications
  Magazine}, vol.~56, no.~2, pp. 219--224, 2018.

\bibitem{RademacherOFC20}
G.~Rademacher, B.~J. Puttnam, R.~S. Luís, J.~Sakaguchi, W.~Klaus, T.~A.
  Eriksson, Y.~Awaji, T.~Hayashi, T.~Nagashima, T.~Nakanishi, T.~Taru,
  T.~Takahata, T.~Kobayashi, H.~Furukawa, and N.~Wada, ``10.66 {Peta-Bit/s}
  transmission over a 38-core-three-mode fiber,'' in \emph{2020 Optical Fiber
  Communications Conference and Exhibition (OFC)}, 2020, pp. 1--3.

\bibitem{pagare2021design}
R.~A. Pagare, S.~Kumar, and A.~Mishra, ``Design and analysis of hybrid optical
  distribution network for worst-case scenario of {E2}-class symmetric
  coexistence 80 {G}bps {TWDM} {NG-PON2} architecture for {FTTX} access
  networks,'' \emph{Optik}, vol. 228, p. 166168, 2021.

\bibitem{zhu20221lambda}
Y.~Zhu, C.~Zhang, X.~Zeng, H.~Jiang, Y.~Xu, X.~Xie, Q.~Zhuge, and W.~Hu,
  ``1$\lambda$ 10.5 {T}b/s {CPRI}-equivalent rate 1024-{QAM} transmission via
  self-homodyne digital-analog radio-over-fiber architecture,'' in
  \emph{European Conference and Exhibition on Optical Communication}.\hskip 1em
  plus 0.5em minus 0.4em\relax Optica Publishing Group, 2022, pp. Th3A--5.

\bibitem{LimJLT21}
C.~Lim and A.~Nirmalathas, ``Radio-over-fiber technology: {Present} and
  future,'' \emph{Journal of Lightwave Technology}, vol.~39, no.~4, pp.
  881--888, Feb 2021.

\bibitem{TrichiliJOSAB20}
A.~Trichili, M.~A. Cox, B.~S. Ooi, and M.-S. Alouini, ``Roadmap to free space
  optics,'' \emph{J. Opt. Soc. Am. B}, vol.~37, pp. A184--A201, 2020.

\bibitem{CiaramellaJSAC09}
E.~Ciaramella, Y.~Arimoto, G.~Contestabile, M.~Presi, A.~D'Errico, V.~Guarino,
  and M.~Matsumoto, ``1.28 {Terabit/s (32x40 Gbit/s) WDM} transmission system
  for free space optical communications,'' \emph{IEEE Journal on Selected Areas
  in Communications}, vol.~27, no.~9, pp. 1639--1645, 2009.

\bibitem{curran2017fsonet}
M.~Curran, M.~S. Rahman, H.~Gupta, K.~Zheng, J.~Longtin, S.~R. Das, and
  T.~Mohamed, ``{FSONet}: A wireless backhaul for multi-gigabit picocells using
  steerable free space optics,'' in \emph{Proceedings of the 23rd Annual
  International Conference on Mobile Computing and Networking}, 2017, pp.
  154--166.

\bibitem{cuervo2019mixed}
E.~Cuervo, M.~Ghobadi, K.~Chintalapudi, and M.~Kotaru, ``Mixed reality offload
  using free space optics,'' Dec.~17 2019, uS Patent 10,509,463.

\bibitem{rahman2018fso}
M.~S. Rahman, K.~Zheng, and H.~Gupta, ``{FSO-VR}: steerable free space optics
  link for virtual reality headsets,'' in \emph{Proceedings of the 4th ACM
  Workshop on Wearable Systems and Applications}, 2018, pp. 11--15.

\bibitem{CaiOFC18}
J.-X. Cai, H.~G. Batshon, M.~V. Mazurczyk, O.~V. Sinkin, D.~Wang, M.~Paskov,
  C.~R. Davidson, W.~W. Patterson, A.~Turukhin, M.~A. Bolshtyansky, and D.~G.
  Foursa, ``51.5 {Tb/s} capacity over 17,107 km in {C+L} bandwidth using
  single-mode fibers and nonlinearity compensation,'' \emph{Journal of
  Lightwave Technology}, vol.~36, no.~11, pp. 2135--2141, 2018.

\bibitem{ShiuPJ20}
R.-K. Shiu, Y.-W. Chen, P.-C. Peng, S.-J. Su, G.-M. Shao, J.~Chiu, J.-W. Li,
  and G.-K. Chang, ``A simplified radio-over-fiber system for over 100-km
  long-reach n-qam transmission,'' \emph{IEEE Photonics Journal}, vol.~12,
  no.~3, pp. 1--8, 2020.

\bibitem{Taara}
``High-throughput links move data wirelessly,''
  \url{https://x.company/projects/taara/}, last accessed 26.09.2022.

\bibitem{FeiOPEX22}
C.~Fei, Y.~Wang, J.~Du, R.~Chen, N.~Lv, G.~Zhang, J.~Tian, X.~Hong, and S.~He,
  ``100-m/3-{Gbps} underwater wireless optical transmission using a wideband
  photomultiplier tube ({PMT}),'' \emph{Optics Express}, vol.~30, no.~2, pp.
  2326--2337, Jan 2022.

\bibitem{BlueComm200}
``Underwater optical communications and data transfer modem,''
  \url{https://www.sonardyne.com/products/bluecomm-200-wireless-underwater-link/},
  last accessed 26.09.2022.

\bibitem{BianJLT19}
R.~Bian, I.~Tavakkolnia, and H.~Haas, ``15.73 {Gb/s} visible light
  communication with off-the-shelf {LEDs},'' \emph{Journal of Lightwave
  Technology}, vol.~37, no.~10, pp. 2418--2424, May 2019.

\bibitem{WangOE16}
Y.~Wang, X.~Huang, J.~Shi, Y.~quan Wang, and N.~Chi, ``{Long-range high-speed
  visible light communication system over 100-m outdoor transmission utilizing
  receiver diversity technology},'' \emph{Optical Engineering}, vol.~55, no.~5,
  p. 056104, 2016.

\bibitem{WeiOPEX19}
L.-Y. Wei, C.-W. Chow, G.-H. Chen, Y.~Liu, C.-H. Yeh, and C.-W. Hsu, ``Tricolor
  visible-light laser diodes based visible light communication operated at
  40.665 {Gbit/s} and 2 m free-space transmission,'' \emph{Optics Express},
  vol.~27, no.~18, pp. 25\,072--25\,077, Sep 2019.

\bibitem{QinACPC21}
G.~Qin, Q.~Bian, W.~Niu, and N.~Chi, ``100m free-space visible light
  communication at {6 Gbps GS-APSK} modulation utilizing a {GaN} blue {LD},''
  in \emph{Asia Communications and Photonics Conference 2021}.\hskip 1em plus
  0.5em minus 0.4em\relax Optica Publishing Group, 2021, p. T4A.69.

\bibitem{LiCrystals22}
D.~Li, C.~Ma, J.~Wang, F.~Hu, Y.~Hou, S.~Wang, J.~Hu, S.~Yi, Y.~Ma, J.~Shi,
  J.~Zhang, Z.~Li, N.~Chi, L.~Xia, and C.~Shen, ``High-speed {GaN}-based
  superluminescent diode for 4.57 {Gbps} visible light communication,''
  \emph{Crystals}, vol.~12, no.~2, 2022.

\bibitem{ChengOPEX22}
C.~Cheng, X.~Li, Q.~Xiang, J.~Li, Y.~Jin, Z.~Wei, H.~Y. Fu, and Y.~Yang,
  ``{4-bit DAC based 6.9Gb/s PAM-8 UOWC system using single-pixel mini-LED and
  digital pre-compensation},'' \emph{Optics Express}, vol.~30, no.~15, pp.
  28\,014--28\,023, Jul 2022.

\bibitem{ShenOptCom19}
J.~Shen, J.~Wang, C.~Yu, X.~Chen, J.~Wu, M.~Zhao, F.~Qu, Z.~Xu, J.~Han, and
  J.~Xu, ``Single led-based 46-m underwater wireless optical communication
  enabled by a multi-pixel photon counter with digital output,'' \emph{Optics
  Communications}, vol. 438, pp. 78--82, 2019.

\bibitem{TsaiSREP19}
W.-S. Tsai, H.-H. Lu, H.-W. Wu, C.-W. Su, and Y.-C. Huang, ``A 30 {Gb/s PAM4}
  underwater wireless laser transmission system with optical beam
  reducer/expander,'' \emph{Scientific Reports}, vol.~9, no.~1, p. 8605, Jun
  2019.

\bibitem{ZhouJLT22}
H.~Zhou, M.~Zhang, X.~Wang, and X.~Ren, ``Design and implementation of more
  than 50m real-time underwater wireless optical communication system,''
  \emph{Journal of Lightwave Technology}, vol.~40, no.~12, pp. 3654--3668,
  2022.

\bibitem{HaasPTRSA20}
H.~Haas, J.~Elmirghani, and I.~White, ``Optical wireless communication,''
  \emph{Philos. Trans. R. Soc. A}, vol. 378, no. 2169, p. 20200051, 2020.

\bibitem{janjua2015going}
B.~Janjua, H.~M. Oubei, J.~R.~D. Retamal, T.~K. Ng, C.-T. Tsai, H.-Y. Wang,
  Y.-C. Chi, H.-C. Kuo, G.-R. Lin, J.-H. He \emph{et~al.}, ``Going beyond 4
  {G}bps data rate by employing {RGB} laser diodes for visible light
  communication,'' \emph{Optics express}, vol.~23, no.~14, pp.
  18\,746--18\,753, 2015.

\bibitem{lee20152}
C.~Lee, C.~Shen, H.~M. Oubei, M.~Cantore, B.~Janjua, T.~K. Ng, R.~M. Farrell,
  M.~M. El-Desouki, J.~S. Speck, S.~Nakamura \emph{et~al.}, ``2 {G}bit/s data
  transmission from an unfiltered laser-based phosphor-converted white lighting
  communication system,'' \emph{Optics express}, vol.~23, no.~23, pp.
  29\,779--29\,787, 2015.

\bibitem{shen2019group}
C.~Shen, J.~A. Holguin-Lerma, A.~A. Alatawi, P.~Zou, N.~Chi, T.~K. Ng, and
  B.~S. Ooi, ``Group-iii-nitride superluminescent diodes for solid-state
  lighting and high-speed visible light communications,'' \emph{IEEE Journal of
  Selected Topics in Quantum Electronics}, vol.~25, no.~6, pp. 1--10, 2019.

\bibitem{alatawi2018high}
A.~A. Alatawi, J.~A. Holguin-Lerma, C.~H. Kang, C.~Shen, R.~C. Subedi, A.~M.
  Albadri, A.~Y. Alyamani, T.~K. Ng, and B.~S. Ooi, ``High-power blue
  superluminescent diode for high {CRI} lighting and high-speed visible light
  communication,'' \emph{Optics Express}, vol.~26, no.~20, pp.
  26\,355--26\,364, 2018.

\bibitem{swimvr}
{swim VR}, ``Swim vr brings the ocean in every pool,'' https://swim-vr.com/.

\bibitem{sinnott2019underwater}
C.~Sinnott, J.~Liu, C.~Matera, S.~Halow, A.~Jones, M.~Moroz, J.~Mulligan,
  M.~Crognale, E.~Folmer, and P.~MacNeilage, ``Underwater virtual reality
  system for neutral buoyancy training: Development and evaluation,'' in
  \emph{Proceedings of the 25th ACM Symposium on Virtual Reality Software and
  Technology}, 2019, pp. 1--9.

\bibitem{yamashita2016aquacave}
S.~Yamashita, X.~Zhang, T.~Miyaki, and J.~Rekimoto, ``Aquacave: An underwater
  immersive projection system for enhancing the swimming experience.'' in
  \emph{ICAT-EGVE}, 2016, pp. 25--28.

\bibitem{snorkel}
{vr snorkel}, ``Vr snorkeling experiences,'' https://vr-snorkel.com/.

\bibitem{fauville2021effect}
G.~Fauville, A.~Queiroz, E.~S. Woolsey, J.~W. Kelly, and J.~N. Bailenson, ``The
  effect of water immersion on vection in virtual reality,'' \emph{Scientific
  Reports}, vol.~11, no.~1, pp. 1--13, 2021.

\bibitem{NG}
{NG}, ``Nemo's garden,'' http://www.nemosgarden.com/.

\bibitem{Siemens}
{Siemens}, ``Siemens shows path to industrial metaverse at web summit 2022,''
  https://press.siemens.com/global/en/pressrelease/siemens-shows-path-industrial-metaverse-web-summit-2022.

\bibitem{PuttnamECOC15}
B.~J. Puttnam, R.~S. Luís, W.~Klaus, J.~Sakaguchi, J.-M. Delgado~Mendinueta,
  Y.~Awaji, N.~Wada, Y.~Tamura, T.~Hayashi, M.~Hirano, and J.~Marciante, ``2.15
  pb/s transmission using a 22 core homogeneous single-mode multi-core fiber
  and wideband optical comb,'' in \emph{2015 European Conference on Optical
  Communication (ECOC)}, 2015, pp. 1--3.

\bibitem{KobayashiOFC17}
T.~Kobayashi, M.~Nakamura, F.~Hamaoka, K.~Shibahara, T.~Mizuno, A.~Sano,
  H.~Kawakami, A.~Isoda, M.~Nagatani, H.~Yamazaki, Y.~Miyamoto, Y.~Amma,
  Y.~Sasaki, K.~Takenaga, K.~Aikawa, K.~Saitoh, Y.~Jung, D.~J. Richardson,
  K.~Pulverer, M.~Bohn, M.~Nooruzzaman, and T.~Morioka, ``{1-Pb/s (32 SDM/46
  WDM/768 Gb/s) C-band dense SDM transmission over 205.6-km of single-mode
  heterogeneous multi-core fiber using 96-Gbaud PDM-16QAM channels},'' in
  \emph{2017 Optical Fiber Communications Conference and Exhibition (OFC)},
  2017, pp. 1--3.

\bibitem{RademacherOFC18}
G.~Rademacher, R.~S. Luis, B.~J. Puttnam, T.~A. Eriksson, E.~Agrell,
  R.~Maruyama, K.~Aikawa, H.~Furukawa, Y.~Awaji, and N.~Wada, ``159 {Tbit/s
  C+L} band transmission over 1045 km 3-mode graded-index few-mode fiber,'' in
  \emph{2018 Optical Fiber Communications Conference and Exposition (OFC)},
  2018, pp. 1--3.

\bibitem{CorcoranNatCom20}
B.~Corcoran, M.~Tan, X.~Xu, A.~Boes, J.~Wu, T.~G. Nguyen, S.~T. Chu, B.~E.
  Little, R.~Morandotti, A.~Mitchell, and D.~J. Moss, ``Ultra-dense optical
  data transmission over standard fibre with a single chip source,''
  \emph{Nature Communications}, vol.~11, no.~1, May 2020.

\bibitem{GaldinoPTL20}
L.~Galdino, A.~Edwards, W.~Yi, E.~Sillekens, Y.~Wakayama, T.~Gerard, W.~S.
  Pelouch, S.~Barnes, T.~Tsuritani, R.~I. Killey, D.~Lavery, and P.~Bayvel,
  ``Optical fibre capacity optimisation via continuous bandwidth amplification
  and geometric shaping,'' \emph{IEEE Photonics Technology Letters}, vol.~32,
  no.~17, pp. 1021--1024, 2020.

\bibitem{PuttnamOFC21}
B.~J. Puttnam, R.~S. Luís, G.~Rademacher, Y.~Awaji, and H.~Furukawa, ``319
  {Tb/s} transmission over 3001 km with {S, C and L} band signals over $>$120nm
  bandwidth in 125 $\mu$m wide 4-core fiber,'' in \emph{2021 Optical Fiber
  Communications Conference and Exhibition (OFC)}, 2021, pp. 1--3.

\bibitem{RetamalOPEX15}
B.~Janjua, H.~M. Oubei, J.~R.~D. Retamal, T.~K. Ng, C.-T. Tsai, H.-Y. Wang,
  Y.-C. Chi, H.-C. Kuo, G.-R. Lin, J.-H. He, and B.~S. Ooi, ``Going beyond 4
  gbps data rate by employing rgb laser diodes for visible light
  communication,'' \emph{Optics Express}, vol.~23, no.~14, pp.
  18\,746--18\,753, Jul 2015.

\bibitem{BianECOC18}
R.~Bian, I.~Tavakkolnia, and H.~Haas, ``10.2 {Gb/s} visible light communication
  with off-the-shelf {LEDs},'' in \emph{2018 European Conference on Optical
  Communication (ECOC)}, 2018, pp. 1--3.

\bibitem{ChunJLT16}
H.~Chun, S.~Rajbhandari, G.~Faulkner, D.~Tsonev, E.~Xie, J.~J.~D. McKendry,
  E.~Gu, M.~D. Dawson, D.~C. O'Brien, and H.~Haas, ``{LED} based wavelength
  division multiplexed {10 Gb/s} visible light communications,'' \emph{Journal
  of Lightwave Technology}, vol.~34, no.~13, pp. 3047--3052, 2016.

\bibitem{GutemaJLT22}
T.~Z. Gutema, H.~Haas, and W.~O. Popoola, ``{WDM} based 10.8 {Gbps} visible
  light communication with probabilistic shaping,'' \emph{Journal of Lightwave
  Technology}, vol.~40, no.~15, pp. 5062--5069, 2022.

\bibitem{OubeiOPEX15}
H.~M. Oubei, J.~R. Duran, B.~Janjua, H.-Y. Wang, C.-T. Tsai, Y.-C. Chi, T.~K.
  Ng, H.-C. Kuo, J.-H. He, M.-S. Alouini, G.-R. Lin, and B.~S. Ooi, ``{4.8
  Gbit/s 16-QAM-OFDM transmission based on compact 450-nm laser for underwater
  wireless optical communication},'' \emph{Optics Express}, vol.~23, no.~18,
  pp. 23\,302--23\,309, Sep 2015.

\bibitem{KongOPEX17}
M.~Kong, W.~Lv, T.~Ali, R.~Sarwar, C.~Yu, Y.~Qiu, F.~Qu, Z.~Xu, J.~Han, and
  J.~Xu, ``{10-m 9.51-Gb/s RGB laser diodes-based WDM underwater wireless
  optical communication},'' \emph{Optics Express}, vol.~25, pp.
  20\,829--20\,834, Aug 2017.

\bibitem{Liu2019}
X.~Liu, ``Evolution of fiber-optic transmission and networking toward the {5G}
  era,'' \emph{{iScience}}, vol.~22, pp. 489--506, Dec. 2019.

\bibitem{chen2015downlink}
C.~Chen, D.~A. Basnayaka, and H.~Haas, ``Downlink performance of optical
  attocell networks,'' \emph{Journal of Lightwave Technology}, vol.~34, no.~1,
  pp. 137--156, 2015.

\bibitem{kuhlmann2023static}
L.~Kuhlmann~de Canaviri, K.~Meiszl, V.~Hussein, P.~Abbassi, S.~D.
  Mirraziroudsari, L.~Hake, T.~Potthast, F.~Ratert, T.~Schulten, M.~Silberbach
  \emph{et~al.}, ``Static and dynamic accuracy and occlusion robustness of
  steamvr tracking 2.0 in multi-base station setups,'' \emph{Sensors}, vol.~23,
  no.~2, p. 725, 2023.

\bibitem{Streamvr}
S.~{Tracking}, ``Sub-millimeter accuracy,''
  https://partner.steamgames.com/vrlicensing.

\bibitem{valve}
{VALVE}, ``Sub-millimeter resolution to capture every gesture,''
  https://www.valvesoftware.com/en/index/base-stations.

\bibitem{IRSLiFi22}
H.~Abumarshoud, L.~Mohjazi, O.~A. Dobre, M.~Di~Renzo, M.~A. Imran, and H.~Haas,
  ``{LiFi} through reconfigurable intelligent surfaces: {A} new frontier for
  {6G}?'' \emph{IEEE Vehicular Technology Magazine}, vol.~17, no.~1, pp.
  37--46, 2022.

\bibitem{cao2019reconfigurable}
Z.~Cao, X.~Zhang, G.~Osnabrugge, J.~Li, I.~M. Vellekoop, and A.~M. Koonen,
  ``Reconfigurable beam system for non-line-of-sight free-space optical
  communication,'' \emph{Light: Science \& Applications}, vol.~8, no.~1, p.~69,
  2019.

\bibitem{poletti2013towards}
F.~e. Poletti, N.~Wheeler, M.~Petrovich, N.~Baddela, E.~Numkam~Fokoua,
  J.~Hayes, D.~Gray, Z.~Li, R.~Slav{\'\i}k, and D.~Richardson, ``Towards
  high-capacity fibre-optic communications at the speed of light in vacuum,''
  \emph{Nature Photonics}, vol.~7, no.~4, pp. 279--284, 2013.

\bibitem{blogHCF}
{Official Microsoft Blog}, ``Microsoft acquires {L}umenisity®, an innovator in
  hollow core fiber ({HCF}) cable,''
  https://blogs.microsoft.com/blog/2022/12/09/microsoft-acquires-lumenisity-an-innovator-in-hollow-core-fiber-hcf-cable/.

\bibitem{de2019photonic}
L.~De~Marinis, M.~Cococcioni, P.~Castoldi, and N.~Andriolli, ``Photonic neural
  networks: A survey,'' \emph{IEEE Access}, vol.~7, pp. 175\,827--175\,841,
  2019.

\end{thebibliography}

\end{document}